# ANTIDS: SELF-ORGANIZED ANT-BASED CLUSTERING MODEL FOR INTRUSION DETECTION SYSTEM

Vitorino Ramos
*CVRM-IST, Technical University of Lisbon (IST)*
*Av. Rovisco Pais, 1, 1049-001, Lisbon, PORTUGAL*
vitorino.ramos@alfa.ist.utl.pt

Ajith Abraham
*School of Computer Science and Engineering, Chung-Ang University*
*211 Heukseok-dong, Dongjak-gu, Seoul 156-756, SOUTH KOREA*
ajith.abraham@ieee.org

**ABSTRACT**

Security of computers and the networks that connect them is increasingly becoming of great significance. Computer security is defined as the protection of computing systems against threats to confidentiality, integrity, and availability. There are two types of intruders: the external intruders who are unauthorized users of the machines they attack, and internal intruders, who have permission to access the system with some restrictions. Due to the fact that it is more and more improbable to a system administrator to recognize and manually intervene to stop an attack, there is an increasing recognition that ID systems should have a lot to earn on following its basic principles on the behavior of complex natural systems, namely in what refers to self-organization, allowing for a real distributed and collective perception of this phenomena. Having that aim in mind, the present work presents a self-organized ant colony based intrusion detection system (ANTIDS) to detect intrusions in a network infrastructure. The performance is compared among conventional soft computing paradigms like Decision Trees, Support Vector Machines and Linear Genetic Programming to model fast, online and efficient intrusion detection systems.

## 1. INTRODUCTION

The process of monitoring the events occurring in a computer system or network and analyzing them for sign of intrusions is known as Intrusion detection. Intrusion detection is classified into two types: misuse intrusion detection and anomaly intrusion detection. Misuse intrusion detection uses well-defined patterns of the attack that exploit weaknesses in system and application software to identify the intrusions. These patterns are encoded in advance and used to match against the user behavior to detect intrusion. Anomaly intrusion detection uses the normal usage behavior patterns to identify the intrusion. The behavior of the user is observed and any deviation from the constructed normal behavior is detected as intrusion. We have two options to secure the system completely, either prevent the threats and vulnerabilities which come from flaws in the operating system as well as in the application programs or detect them and take some action to prevent them in future and also repair the damage. It is impossible in practice, and even if possible, extremely difficult and expensive, to write a completely secure system. Transition to such a system for use in the entire world would be an equally difficult task. Cryptographic methods can be compromised if the passwords and keys are stolen. No matter how secure a system is, it is vulnerable to insiders who abuse their privileges. There is an inverse relationship between the level of access control and efficiency. More access controls make a system less user-friendly and more likely of not being used.

An Intrusion Detection System (IDS) is a program that analyzes what happens or has happened during an execution and tries to find indications that the computer has been misused. An Intrusion detection system does not eliminate the use of preventive mechanism but it works as the last defensive mechanism in securing the system. Data mining approaches for intrusion detection were first implemented in mining audit data for automated models for intrusion detection [3,23,8].. Several data mining algorithms are applied to audit data to compute models that accurately capture the actual behavior of intrusions as well as normal activities [29]. Audit data analysis and mining combine the association rules and classification algorithm to discover attacks in audit data.

On the other hand, and due to the fact that it is more and more improbable to a system administrator to recognize and manually intervene to stop an attack (an option only possible in small scale networks) without harming too much the integrity of the overall system, there is an increasing recognition that ID systems should have a lot to earn on following its basic principles on the behavior of complex natural systems, namely in what refers to self-organization. Due to their nature, self-organizing complex adaptive systems typically are comprised of a large number of frequently similar components (e.g. agents) or events. Through their process, a pattern at the global-level of a system emerges solely from numerous interactions among the lower-level components [31,35]. Moreover, the rules specifying interactions among the system's components are executed using only local information, without reference to the global pattern, which, as in many real-world problems is not easily accessible or possible to be found, as in our IDS present case. Stigmergy [34,33,2], a kind of indirect communication and learning by the environment found in social insects is a well know example of self-organization, providing not only vital clues in order to understand how the components can interact to produce a complex pattern, as can pinpoint simple biological non-linear rules and methods to achieve improved artificial intelligent adaptive categorization systems, critical for collective perception and recognition. In fact, their distributed bottom-up emergent nature, along with their massively implicit parallel properties and the fact that there is no need of a global top-down hierarchical supervisor, makes them ideal candidates for ID systems and to be embedded on-line in complex large scale computer network infrastructures, where traditional security mechanisms demonstrate severe weaknesses [25]. Some works have already been presented along these lines. Recently, *Foukia* et al [16] designed an IR (Intrusion Response) system cooperating with an IDS using mobile agents distributed throughout the network, based on stigmergic properties. In other works [14,11], detection was based on artificial immune systems [10] where ID agents map the functionalities of the natural immune system to distinguish between normal and abnormal events (respectively "self" and "non self" in the immune system) as explained in [17,18].

Precisely on the context of adopting complex adaptive systems into ID systems, the present work introduces a self-organized ant colony based intrusion detection system (ANTIDS) to detect intrusions and compares its performance with Linear Genetic Programming (LGP) [5], Support Vector Machines (SVM) [38] and Decision Trees (DT) [6]. Other past works have made use of Multiple Adaptive Regression Splines (MARS) [28]. The rest of the paper is organized as follows. Section 2 presents some related research in intrusion detection systems using data mining paradigms. The technical details of the ANT colony algorithm are presented in Section 3 followed by the importance of attribute or feature reduction in Section 4. Experiment results are presented in Section 5 and some conclusions are also provided towards the end.

## 2. INTRUSION DETECTION SYSTEM – A DATA MINING APPROACH

Data mining is a relatively new approach for intrusion detection. Data mining approaches for intrusion detection was first implemented in Mining Audit Data for Automated Models for Intrusion Detection [26]. The raw data is first converted into ASCII network packet information which in turn is converted into connection level information. These connection level records contain within connection features like service, duration, etc. Data mining algorithms are applied to this data to create models to detect intrusions.

### 2.1 Linear Genetic Programming (LGP)

Linear genetic programming is a variant of the GP technique that acts on linear genomes. Its main characteristics in comparison to tree-based GP lies in that the evolvable units are not the expressions of a functional programming language (like LISP), but the programs of an imperative language (like c/c ++) [5]. The basic unit of evolution here is a native machine code instruction that runs on the floating-point processor unit (FPU). Since different instructions may have different sizes, here instructions are clubbed up together to form *instruction blocks* of 32 bits each. The *instruction blocks* hold one or more native machine code instructions, depending on the sizes of the instructions. A crossover point can occur only between instructions and is prohibited from occurring within an instruction. However the mutation operation does not have any such restriction. In this research, steady state genetic programming approach was used to manage the memory more effectively.

### 2.2 Decision Trees (DT)

Intrusion detection can be considered as classification problem where each connection or user is identified either as one of the attack types or normal based on some existing data. Decision trees work well with large data sets. This is important as large amounts of data flow across computer networks. The high performance of Decision trees makes them useful in real-time intrusion detection. Decision trees construct easily interpretable models, which is useful for a security officer to inspect and edit. These models can also be used in the rule-based models with minimum processing [6]. Generalization accuracy of decision trees is another useful property for intrusion detection model. There will always be some new attacks on the system, which are small variations of known attacks after the intrusion detection models are built. The ability to detect these new intrusions is possible due to the generalization accuracy of decision trees.

### 2.3 Support Vector Machines (SVM)

Support Vector Machines have been proposed as a novel technique for intrusion detection. SVM maps input (real-valued) feature vectors into a higher dimensional feature space through some nonlinear mapping. SVMs are powerful tools for providing solutions to classification, regression and density estimation problems. These are developed on the principle of structural risk minimization. Structural risk minimization seeks to find a hypothesis h for which one can find lowest probability of error. The structural risk minimization can be achieved by finding the hyper plane with maximum separable margin for the data [38]. Computing the hyper plane to separate the data points i.e. training a SVM leads to quadratic optimization problem. SVM uses a feature called kernel to solve this problem. Kernel transforms linear algorithms into nonlinear ones via a map into feature spaces. SVMs classify data by using these support vectors, which are members of the set of training inputs that outline a hyper plane in feature space.

## 3. BIO-INSPIRED SELF-ORGANIZED ANT-BASED CLUSTERING

Data Mining is precisely one of those problems in which real ants can suggest very interesting heuristics for computer scientists [31], namely for clustering purposes. One of the first studies using the metaphor of ant colonies related to the above clustering domain is due to *Deneubourg* [9], where a population of ant-like agents randomly moving onto a 2D grid are allowed to move basic objects so as to cluster them. This method was then further generalized by *Lumer* and *Faieta* [24] (here after *LF* algorithm), applying it to exploratory data analysis, for the first time. In 1995, the two authors were then beyond the simple example, and applied their algorithm to interactive exploratory database analysis, where a human observer can probe the contents of each represented point (sample, image, item) and alter the characteristics of the clusters. They showed that their model provides a way of exploring complex information spaces, such as document or relational databases, because it allows information access based on exploration from various perspectives. However, this last work entitled "Exploratory Database Analysis via Self-Organization", according to [4], was never published due to commercial applications. They applied the algorithm to a database containing the "profiles" of 1650 bank customers. Attributes of the profiles included marital status, gender, residential status, age, a list of banking services used by the customer, etc. Given the variety of attributes, some of them qualitative and others quantitative, they had to define several dissimilarity measures for the different classes of attributes, and to combine them into a global dissimilarity measure (in, pp. 163, Chapter 4 [4]). More recently, *Ramos* et al. [34,33,35] presented a novel strategy (*ACLUSTER*) to tackle unsupervised clustering as well as data retrieval problems, avoiding not only short-term memory based strategies, as well as the use of several artificial ant types (using different speeds), present in those approaches proposed initially by *Lumer* [24]. Other works in this area include those from *Monmarché* et al. [27], *Ramos*, *Merelo* et al. [35,33,34,31], *Handl* and *Dorigo* [19], *Ramos* and *Abraham* [32,2].

**Algorithm.** High-level description of *ACLUSTER*.

```
/* Initialization */
For every object or data-item o_i do
Place o_i randomly on grid
End For
For all agents do
Place agent at randomly selected site
End For

/* Main loop */
For t = 1 to t_max do
For all agents do
sum = 0
Count the number of items n around site r

If ((agent unladen) & (site r occupied by item o_i)) then
For all sites around r with items present do
/* According to Eqs. 4, 6, 7 and 8 */
Compute d,χ, e and P_p
Draw a random real number R between 0 and 1
If (R = P_p) then
sum = sum + 1
End If
End For
If ((sum = n/2) or (n = 0)) then
Pick up item o_i
End If
Else If ((agent carrying item o_i) & (site r empty)) then
For all sites around r with items present do
/* According to Eqs. 4, 5, 7 and 8 */
Compute d,χ, d and P_d
Draw random real number R between 0 and 1
If (R = P_d) then
sum = sum + 1
End If
End For
If (sum = n/2) then
Drop item o_i
End If
End If

/* According to Eqs. 1 and 2 (section 3.1) */
Compute W(s) and P_ik
Move to a selected r not occupied by other agent
Count the number of items n around that new site r
Increase pheromone at site r according to n, that is:
P_r = P_r + [h+(n/a)]
End For
Evaporate pheromone by K, at all grid sites
End For
Print location of items

/* Values of parameters used in experiments */
k_1 = 0.1, k_2 = 0.3, K = 0.015, h = 0.07, a = 400,
b = 3.5, ?=0.2, t_max = 10^6 steps.
```

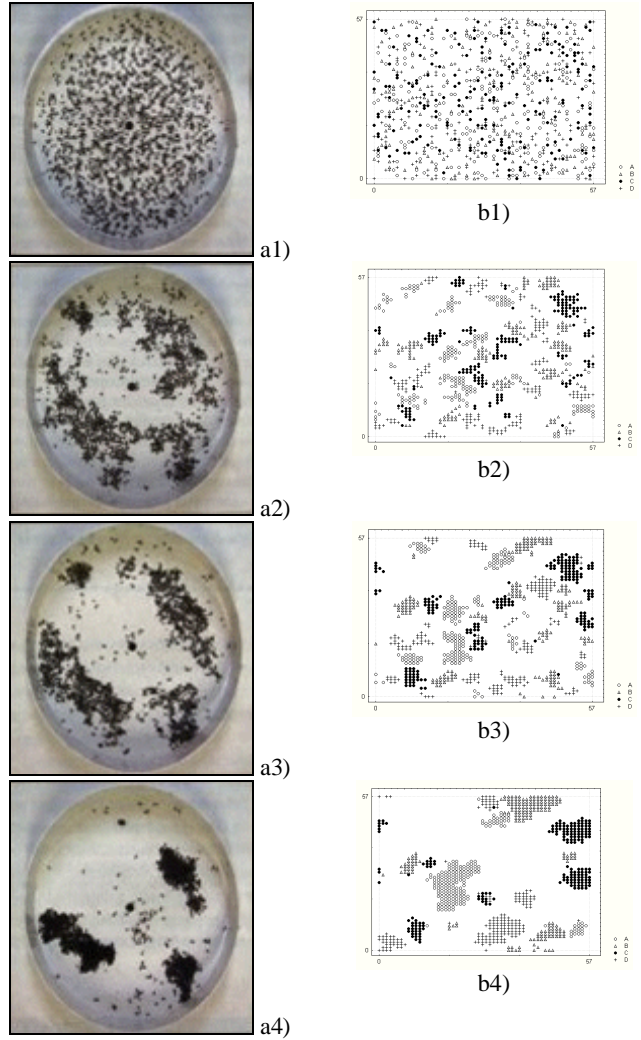

Fig. 1 – From a1) to a4), a sequential clustering task of corpses performed by a real ant colony. 1500 corpses are randomly located in a circular arena with radius = 25 cm, where *Messor Sancta* ant workers are present. The figure shows the initial state (a1), 2 hours (a2), 6 hours (a3) and 26 hours (a4) after the beginning of the experiment [4]. In b1-b4, some experiments with the present algorithm, conducted on synthetic data (as in [33,24]). Spatial distribution of 800 items on a 57 x 57 non-parametric toroidal grid at several time steps. At *t*=1, four types of items are randomly allocated into the grid. As time evolves, several homogenous clusters emerge due to the ant colony action, and as expected the global entropy decreases [34].

## 3.1 Distributed, Collaborative and Stigmergic Clustering

The swarm intelligence algorithm (see fig. 1) fully uses agents that stochastically move around the classification "habitat" following pheromone concentrations. That is, instead of trying to solve some disparities in the basic *LF* algorithm by adding different ant casts, short-term memories and behavioral switches, which are computationally intensive, representing simultaneously a potential and difficult complex parameter tuning, it was our intention to follow real ant-like behaviors as possible (some other features will be incorporated, as the use of different response thresholds to task-associated stimulus intensities, discussed

$$W(s) = \left(1 + \frac{s}{1+ds}\right)^b \quad (1) \qquad P_{ik} = \frac{W(s_i)w(\Delta_i)}{\sum_{j/k} W(s_j)w(\Delta_j)} \quad (2)$$

later). In that sense, bio-inspired spatial transition probabilities are incorporated into the system, avoiding randomly moving agents, which tend the distributed algorithm to explore regions manifestly without interest (e.g., regions without any type of object clusters), being generally, this type of exploration, counterproductive and time consuming. Since this type of transition probabilities depend on the spatial distribution of pheromone across the environment, the behavior reproduced is also a stigmergic one [34,9]. Moreover, the strategy not only allows to guide ants to find clusters of objects in an adaptive way (if, by any reason, one cluster disappears, pheromone tends to evaporate on that location), as the use of embodied short-term memories is avoided (since this transition probabilities tends also to increase pheromone in specific locations, where more objects are present). As we shall see, the distribution of the pheromone represents the memory of the recent history of the swarm, and in a sense it contains information which the individual ants are unable to hold or transmit. There is no direct communication between the organisms but a type of indirect communication through the pheromonal field. In fact, ants are not allowed to have any memory and the individual's spatial knowledge is restricted to local information about the whole colony pheromone density. In order to design this behavior, one simple model was adopted [7], and extended (as in [35,33]) due to specific constraints of the present proposal. As described in [7], the state of an individual ant can be expressed by its position $r$, and orientation $q$. It is then sufficient to specify a transition probability from one place and orientation $(r,q)$ to the next $(r^*,q^*)$ an instant later. The response function can effectively be translated into a two-parameter transition rule between the cells by use of a pheromone weighting function (Eq. 1). This equation measures the relative probabilities of moving to a cite $r$ (in our context, to a grid location) with pheromone density $s(r)$. The parameter $b$ is associated with the osmotropotaxic sensitivity (a kind of instantaneous pheromonal gradient following), and on the other hand, $1/d$ is the sensory capacity, which describes the fact that each ant's ability to sense pheromone decreases somewhat at high concentrations. In addition to the former equation, there is a weighting factor $w(Dq)$, where $Dq$ is the change in direction at each time step, i.e. measures the magnitude of the difference in orientation. As an additional condition, each individual leaves a constant amount $h$ of pheromone at the cell in which it is located at every time step $t$. This pheromone decays at each time step at a rate $k$. Then, the normalised transition probabilities on the lattice to go from cell $k$ to cell $i$ are given by $P_{ik}$ [7] (Eq. 2), where the notation $j/k$ indicates the sum over all the pixels $j$ which are in the local neighborhood of $k$. Finally, $D_i$ measures the magnitude of the difference in orientation for the previous direction at time $t$-1.

## 3.2 Picking and Dropping Data Objects

In order to model the behavior of ants associated to different tasks, as dropping and picking up objects, other works [33] suggest the use of combinations of different response thresholds. As we have seen before, there are two major factors that should influence any local action taken by the ant-like agent: the number of objects in his neighborhood, and their similarity (including the hypothetical object carried by one ant). *Lumer* and *Faieta* [24], use an average similarity, mixing distances between objects with their number, incorporating it simultaneously into a response threshold function. Instead, we recommend the use of combinations of two independent response threshold functions, each associated with a different environmental factor (or, stimuli intensity), that is, the number of objects in the area, and their similarity.

Moreover, the computation of average similarities are avoided in the present algorithm, since this strategy can be somehow blind to the number of objects present in one specific neighborhood. In fact, in *Lumer* and *Faieta*'s work [24], there is an hypothetical chance of having the same average similarity value, respectively having one or, more objects present in that region. But, experimental evidences and observation in some types of ant colonies can provide us with a different answer. After *Wilson* (*The Insect Societies*, Cambridge Press, 1971), it is known that minors and majors in the polymorphic species of ants *Genus Pheidole*, have different response thresholds to task-associated stimulus intensities (i.e., division of labor). Recently, and inspired by this experimental evidence, *Bonabeau* et al. [4], proposed a family of response threshold functions in order to model this behavior. According to it, every individual has a response threshold $q$ for

$$T_q(s) = \frac{s^n}{s^n + q^n} \quad (3) \qquad\qquad c = \frac{n^2}{n^2 + 5^2} \quad (4)$$

$$d = \left(\frac{k_1}{k_1 + d}\right)^2 \quad (5) \qquad\qquad e = \left(\frac{d}{k_2 + d}\right)^2 \quad (6)$$

$$P_p = (1-c).e \quad (7) \qquad\qquad P_d = c.d \quad (8)$$

every task. Individuals engage in task performance when the level of the task-associated stimuli *s*, exceeds their thresholds. Author's defined *s* as the intensity of a stimulus associated with a particular task, i.e. *s* can be a number of encounters, a chemical concentration, or any quantitative cue sensed by individuals. One family of response functions $T_q(s)$ (the probability of performing the task as a function of stimulus intensity *s*), that satisfy this requirement is given by (Eq. 3) [4], where *n*>1 determines the steepness of the threshold (normally *n*=2, but similar results can be obtained with other values of *n*>1). Now, at *s* = *q*, this probability is exactly ½. Therefore, individuals with a lower value of *q* are likely to respond to a lower level of stimulus. In order to take account on the number of objects present in one neighborhood, Eq. 13, was used (where, *n* now stands for the number of objects present in one neighborhood, and *q* = 5), defining χ (Eq. 4) as the response threshold associated to the number of items present in a 3 x 3 region around *r* (one specific grid location). Now, in order to take account on the hypothetical similarity between objects, and in each ant action due to this factor, a *Euclidean* normalized distance *d* is computed within all the pairs of objects present in that 3 x 3 region around *r*. Being *a* and *b*, a pair of objects, and $f_a(i)$, $f_b(i)$ their respective feature vectors (being each object defined by *F* features), then $d = (1/d_{max}).[(1/F).\sum_{i=1,F}(f_a(i)-f_b(i))^2]^{1/2}$. Clearly, this distance *d* reaches its maximum (=1, since *d* is normalized by $d_{max}$) when two objects are maximally different, and *d*=0 when they are equally defined by the same *F* features. Moreover, *d* and *e* (Eqs. 5,6), are respectively defined as the response threshold functions associated to the similarity of objects, in case of dropping an object (Eq. 5), and picking it up (Eq. 6), at site *r*. Finally, in every action taken by an agent, and in order to deal, and represent different stimulus intensities (number of items and their similarity), present at each site in the environment visited by one ant, the strategy uses a composition of the above defined response threshold functions (Eqs. 4,5 and 6). Several composed probabilities were analyzed [33] and used as test functions in one preliminary test. The best results were achieved with the test function #1 below (Eqs. 7,8), achieving a high classification rate (out of 4 different functions were used, as well the *LF* algorithm [24]; for comparison reasons – see [34,33]). Alternatively, the system can also be robust feeding the data continuously (for instance, as they appear) as proved in past works [32]. For other algorithm details please consult [33,34,35].

## 4. ATTRIBUTE REDUCTION IN INTRUSION DETECTION SYSTEMS

Complex relationships exist between features, which are difficult for humans to discover. IDS must therefore reduce the amount of data to be processed. This is very important if real-time detection is desired. The easiest way to do this is by doing an intelligent input feature selection. Certain features may contain false correlations, which hinder the process of detecting intrusions. Further, some features may be redundant since the information they add is contained in other features. Extra features can increase computation time, and can impact the accuracy of IDS. Feature selection improves classification by searching for the subset of features, which best classifies the training data.

Feature selection is done based on the contribution the input variables made to the construction of the decision tree. Feature importance is determined by the role of each input variable either as a main splitter or as a surrogate. Surrogate splitters are defined as back-up rules that closely mimic the action of primary splitting rules. Suppose that, in a given model, the algorithm splits data according to variable '*protocol_type*' and if a value for '*protocol_type*' is not available, the algorithm might substitute 'flag' as a good surrogate. Variable importance, for a particular variable is the sum across all nodes in the tree of the improvement scores that the predictor has when it acts as a primary or surrogate (but not competitor) splitter. Example, for node $i$, if the predictor appears as the primary splitter then its contribution towards importance could be given as $i_{importance}$. But if the variable appears as the $n^{th}$ surrogate instead of the primary variable, then the importance becomes $i_{importance} = (p^n) * i_{improvement}$ in which $p$ is the 'surrogate improvement weight' which is a user controlled parameter set between (0-1).

In order to reduce the number of features or pinpoint those that are relevant, other methods could of course be used, as Principal Component Analysis (PCA, [20]). Other possibility, is not only to try to reduce the number of features as at the same time, trying to increase the recognition rate, which conducts us to a multi-modal optimization problem. A successful method using Genetic Algorithms (Evolutionary Computation) was designed earlier by *Ramos* and *Pina* [36], in order to classify 187 images (each described by 117 Mathematical Morphology [36] features) from 14 classical types of Portuguese granites. Their method not only increased the recognition rate to 100% as features were reduced from 117 to 3, representing simultaneously a robust strategy in order to understand the proper nature of the images treated, and their discriminant features.

## 5. EXPERIMENT SETUP AND RESULTS

The data for our experiments was prepared by the 1998 DARPA intrusion detection evaluation program by MIT Lincoln Labs [26]. The LAN was operated like a real environment, but was blasted with multiple attacks. For each TCP/IP connection, 41 various quantitative and qualitative features were extracted. The data set has 41 attributes or features for each connection record plus one class label as given in table 1. The data set contains 24 attack types that could be classified into four main categories Attack types fall into four main categories: **DoS: Denial of Service -** Denial of Service (DoS) is a class of attack where an attacker makes a computing or memory resource too busy or too full to handle legitimate requests, thus denying legitimate users access to a machine; **R2L: Unauthorized Access from a Remote Machine -** A remote to user (R2L) attack is a class of attack where an attacker sends packets to a machine over a network, then exploits the machine's vulnerability to illegally gain local access as a user; **U2Su: Unauthorized Access to Local Super User (root) -** User to root (U2Su) exploits are a class of attacks where an attacker starts out with access to a normal user account on the system and is able to exploit vulnerability to gain root access to the system; **Probing: Surveillance and Other Probing -** Probing is a class of attack where an attacker scans a network to gather information or find known vulnerabilities. An attacker with a map of machines and services that are available on a network can use the information to look for exploits.

Our experiments had three conventional phases namely input feature reduction, training phase and testing phase, namely for the conventional soft-computing paradigms DT, SVM and LGP. An exception is done for ANTIDS as we will see in the next paragraph. In the data reduction phase, important variables for real-time intrusion detection are selected by feature selection. In the training phase, the different soft computing models were constructed using the training data to give maximum generalization accuracy on the unseen data. The test data is then passed through the saved trained model to detect intrusions in the testing phase. The 41 features are labeled as shown in table 1 and the class label is named as *AP*. This data set has five different classes namely *Normal, DoS, R2L, U2R* and *Probes*. The training and test comprises of 5092 and 6890 records respectively [22]. Using all 41 variables could result in a big IDS model, which could be an overhead for online detection. All the training data were scaled to (0-1). The decision tree approach described in Section 4 helped us to reduce the number of variables to 12 variables. The list of reduced variables is illustrated in table 2. Using the original and reduced data sets, we performed a 5-class classification. The normal data belongs to class 1, probe belongs to class 2, denial of service belongs to class 3, user to super user belongs to class 4, remote to local belongs to class 5. All the IDS models are trained and tested with the same set of data.

The settings of various linear genetic programming system parameters are of utmost importance for successful performance of the system [1]. The population size was set at 120,000 and a tournament size of 8 is used for all the 5 classes. Crossover and mutation probability is set at 65-75% and 75-86% respectively for the different classes. Our trial experiments with SVM revealed that the polynomial kernel option often performs well on most of the datasets. We also constructed decision trees using the training data and then testing data was passed through the constructed classifier to classify the attacks [30]. The ANTIDS experimental setup took however a different path, since it is in his underneath essence an unsupervised process. After feature extraction and reduction (section 4), those values described in table 1 were subjected to normalization. This normalization is based on the computation of *max* and *min* values for each of the 41 features. Then, each feature value (over 11982 samples) are normalized into the interval [0, 1], in order to be randomly introduced in the algorithm described in section 3. In our case, each object (each ID sample) manipulated by the artificial ant colony is represent by a feature vector composed of 41 elements or 12 elements, depending if we instead use the reduced data set (table 2).

**Table 1.** Variables for intrusion detection data set

| Variable No. | Variable name | Variable type | Variable label |
|---|---|---|---|
| 1 | duration | continuous | A |
| 2 | protocol_type | discrete | B |
| 3 | service | discrete | C |
| 4 | flag | discrete | D |
| 5 | src_bytes | continuous | E |
| 6 | dst_bytes | continuous | F |
| 7 | land | discrete | G |
| 8 | wrong_fragment | continuous | H |
| 9 | urgent | continuous | I |
| 10 | hot | continuous | J |
| 11 | num_failed_logins | continuous | K |
| 12 | logged_in | discrete | L |
| 13 | num_compromised | continuous | M |
| 14 | root_shell | continuous | N |
| 15 | su_attempted | continuous | O |
| 16 | num_root | continuous | P |
| 17 | num_file_creations | continuous | Q |
| 18 | num_shells | continuous | R |
| 19 | num_access_files | continuous | S |
| 20 | num_outbound_cmds | continuous | T |
| 21 | is_host_login | discrete | U |
| 22 | is_guest_login | discrete | V |
| 23 | count | continuous | W |
| 24 | srv_count | continuous | X |
| 25 | serror_rate | continuous | Y |
| 26 | srv_serror_rate | continuous | X |
| 27 | rerror_rate | continuous | AA |
| 28 | srv_rerror_rate | continuous | AB |
| 29 | same_srv_rate | continuous | AC |
| 30 | diff_srv_rate | continuous | AD |
| 31 | srv_diff_host_rate | continuous | AE |
| 32 | dst_host_count | continuous | AF |
| 33 | dst_host_srv_count | continuous | AG |
| 34 | dst_host_same_srv_rate | continuous | AH |
| 35 | dst_host_diff_srv_rate | continuous | AI |
| 36 | dst_host_same_src_port_rate | continuous | AJ |
| 37 | dst_host_srv_diff_host_rate | continuous | AK |
| 38 | dst_host_serror_rate | continuous | AL |
| 39 | dst_host_srv_serror_rate | continuous | AM |
| 40 | dst_host_rerror_rate | continuous | AN |
| 41 | dst_host_srv_rerror_rate | continuous | AO |

**Table 2.** Reduced variable set

| C, E, F, L, W, X, Y, AB, AE, AF, AG, AI |
|---|

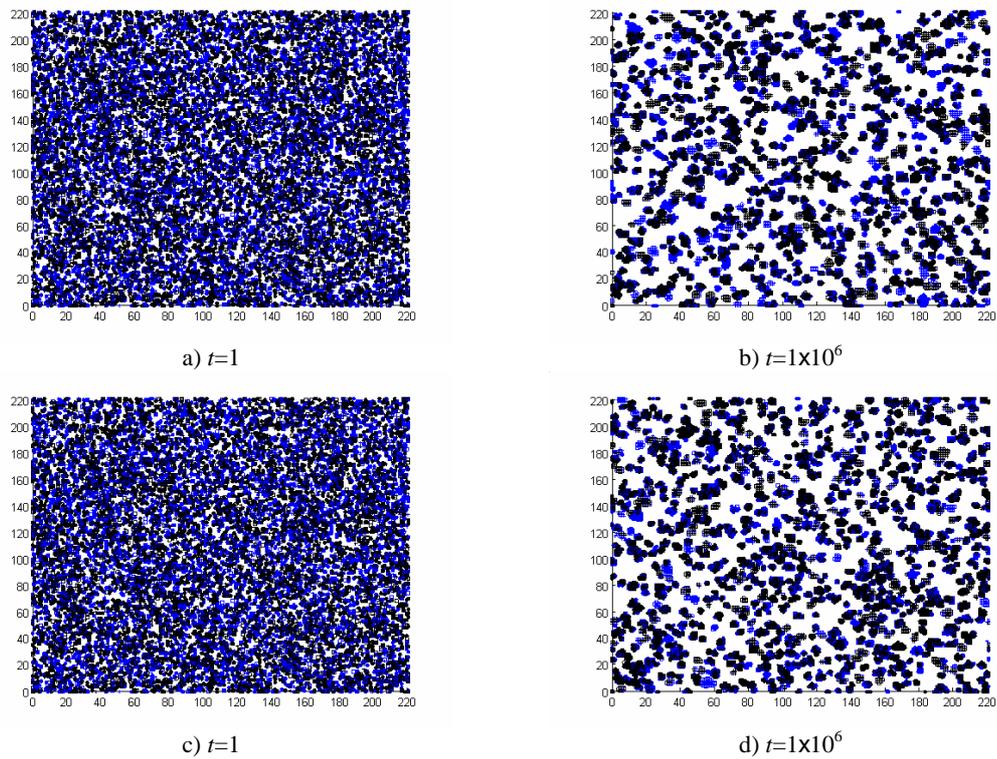

Fig.2 – Ant-like clustering results on IDS data using the a-b) full data set (with 41 features) and c-d) the reduced data set (12 features), at $t=1$ and $t=1 \times 10^6$. For the purpose of reader transparency, in all diagrams class 1 to 5 was respectively represented by ■, □, ●, ○, and finally by ✚ (training samples in blue; testing samples in black).

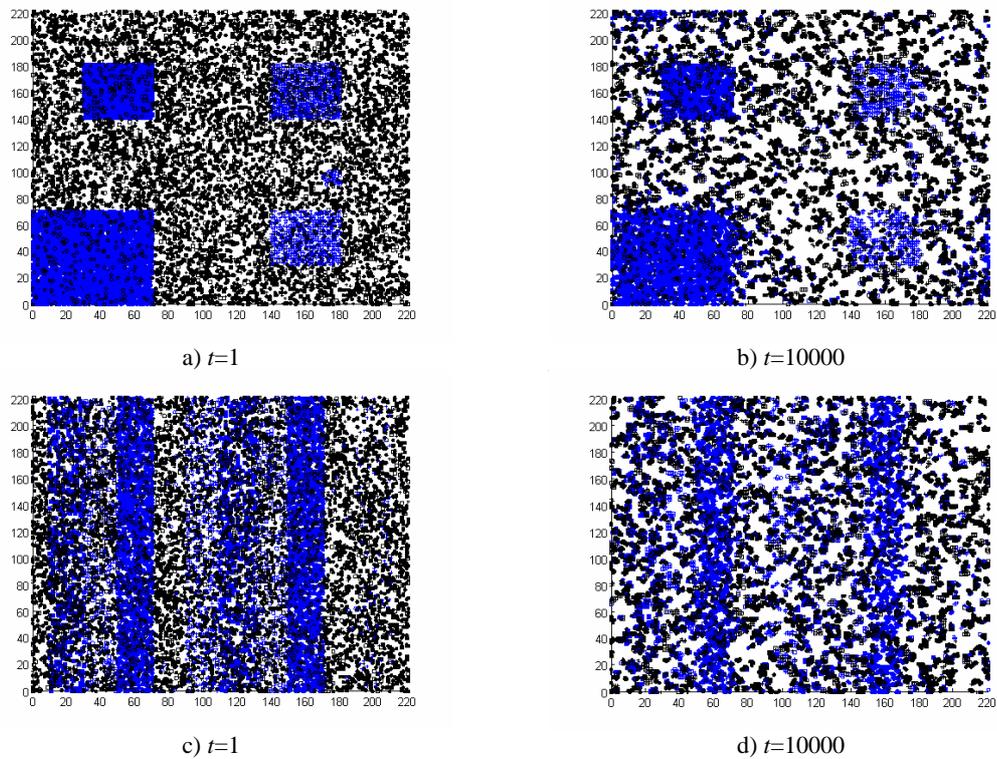

Fig.3 – Other experiments included the initial random allocation of marker items into specified zones.

Based on self-organizing these objects in a non-parametrical toroidal 2-D space, the unsupervised clustering proceeds for $t = 1 \times 10^6$ time steps. After the unsupervised and self-organized clustering process is finished (fig. 2) - with 11982 samples x 41 (12) features each –, the first 5092 samples are used as markers or reference points (blue markers in fig. 2), and via *k*-NNR nearest neighbor rule classification [11,13] the remaining 6890 samples (black markers in fig. 2) are classified (we used $k = 3$ neighbors; *k* must be always an odd number). In order to do so, for each sample $i = 6890, …,11982$, we computed their first $k = 3$ marker neighbors on this non-parametric toroidal 2-D space. An algorithm to find the first *k* neighbors in a toroidal space can however be largely tricky. Our idea was to use 8 virtual spaces (windows) around the one we see in fig. 2, copying to each one of them, all the respective reference markers to be used in the normal *k*-NNR and finally computing the geographical vicinity in this large virtual space for each one of the testing samples only present in the central window. The majority of those marker label values - considered for each still unclassified sample - give them the respective final classification result.

Two experimental setups were then tested. The first one (ANTIDS-a) used self-organization in order to cluster all the 11982 samples at once (5092 training + 6890 testing samples). In ANTIDS-b, however, we process and treat parts and streams of data independently, set after set. In this framework, we use always the full markers set (5092 training samples) plus only a part of the testing set (1000 samples each time). That is, we had to make six runs with 1000 testing samples and the 7[th] final one with 890 testing samples. Final results are presented in tables 3 and 4. For any of these cases, the final self-organized stigmergic map [34] achieved is highly robust. If for example, we use $k = 1$ in the finally *k*-NNR classification, which normally is not prudent, we also arrive at similar recognition rates.

**Table 3.** Performance comparison using full data set

| Attack type | Classification accuracy on test data set (%) | | | | |
|---|---|---|---|---|---|
| | ANTIDS-*a* | ANTIDS-*b* | DT | SVM | LGP |
| Normal | 70.52 | 99.64 | 99.64 | 99.64 | **99.73** |
| Probe | 71.73 | 98.29 | 99.86 | 98.57 | **99.89** |
| DOS | 83.39 | **99.98** | 96.83 | 99.92 | 99.95 |
| U2R | 0.00 | 64.00 | **68.00** | 40.00 | 64.00 |
| R2L | 10.47 | **99.47** | 84.19 | 33.92 | **99.47** |

**Table 4.** Performance comparison using reduced data set

| Attack type | Classification accuracy on test data set (%) | | | | |
|---|---|---|---|---|---|
| | ANTIDS-*a* | ANTIDS-*b* | DT | SVM | LGP |
| Normal | 69.40 | 99.71 | **100.00** | 99.75 | 99.97 |
| Probe | 60.07 | 99.86 | 97.71 | 98.20 | **99.93** |
| DOS | 84.31 | **99.98** | 85.34 | 98.89 | 99.96 |
| U2R | 47.62 | 68.00 | 64.00 | 59.00 | **68.26** |
| R2L | 87.63 | 99.47 | 95.56 | 56.00 | **99.98** |

Other experiments included the initial random allocation of marker items into specified zones (fig. 3). In a) marker samples (training data set for the subsequent *k*-NNR classification; in blue) are randomly allocated in 5 box zones corresponding to samples from class 1 to 5 (class 1 in the bottom-left corner; other classes distributed clockwise), while testing samples are allocated everywhere in this toroidal classification space. These marker samples, however, can now be translated to new places if any ant-like agent wishes to do so, in order to proceed the unsupervised clustering. In c) marker samples are randomly allocated into 10 vertical stripes (2 for each class). In b-d) corresponding results at $t=10000$. Both strategies however, conducts us to meagre recognition rates in the interval [40-72%], for each class, after finally using $k$-NNR ($k=1,3$) at $t=1\times10^6$. For the purpose of reader transparency, in all diagrams class 1 to 5 was respectively represented by ■, □, ●, ○, and finally by ✚ (training samples in blue; testing samples in black). For comparison purposes various other results were adapted from [1,30]. A number of observations and conclusions are drawn from the results illustrated in tables 3 and 4. Using 12 attributes most of the classifiers performed very well.

# 6. CONCLUSION

As seen from results presented in tables 3 and 4 (section 5), the ANTIDS approach has several limitations in what refers to the final recognition rate, obtaining optimal results only for some cases. This is in part due to several reasons. First, the large number of samples used in the present study (a full data set of 11982 items) forces *ACLUSTER* algorithm to be run on a large toroidal space – empirical studies [35,34] shows that the optimal classification "habitat" area should be in the order of 4 times the number of objects, while the number of ants should be in the order of 1/10 of the number of objects. Rather, it's by large preferably to process and treat parts and streams of data independently, set after set (check ANTIDS-b results). Second, and still in the present case, the data is poorly uniformly distributed between all the five classes. In fact, while some classes like class 3 (DOS) are represented by a sum of 3000 training and 4202 testing samples, a total of 7202 items (around 60% of our entire data set), other classes like class 4 (USR) are merely represented by a sum of 27 training and 25 testing samples, a total of 52 items (0.4% of our entire data set). In fact, the probability of one ant to encounter a class 3 sample in the toroidal classification space, process and treat it, is 120 times bigger than to find one from class 4. This fact *per si*, can bias a lot our final results, since any self-organizing mechanism depends a lot on a self-sufficient critical mass. Poorly represented classes have inadequately chances to be classified in a properly manner. It's not surprising at all, that class 4 has the worst results for all methods used for comparison purposes.

However, the self-organizing ID system has 4 major advantages in what refers to a comparison to their counterpart paradigms (DT, SVM, LGP), namely: **(1)** Classification can be processed online and in real time due to their distributed nature, as proved before with swarms nourished with continuous streams of data [32]. In fact, DT, SVM and LGP, as well as Self-Organizing Maps (SOM), have the inability to perform classification and visualization in a continuous basis or to self-organize new data-items into the older ones (even more into new labels if necessary), unless they re-start their learning process from the beginning, a characteristic which can be crucial to our purpose, producing class decisions on a continuous stream data, allowing for continuous mappings with open-ended innovation; **(2)** ANTIDS can deal with new classes whenever it's needed without the need of retraining. In fact, stigmergy is often associated with flexibility: when the environment changes because of an external perturbation, the insects respond *appropriately* to that perturbation, as if it were a modification of the environment caused by the colony's activities. In other words, the colony can collectively respond to the perturbation with individuals exhibiting the same behavior. When it comes to artificial agents, this type of flexibility is priceless: it means that the agents can respond to a perturbation without being reprogrammed to deal with that particular instability. In our context, this means that no classifier re-training is needed for any new sets of data-item types (new classes) arriving to the system, as is necessary in the DT, SVM and LGP models. Moreover, the data-items that were used for comparison purposes in early stages in the colony evolution in his exploration of the search-space, can now, along with new items, be re-arranged in more optimal ways. Classification and/or data retrieval remains the same, but the system re-organizes itself in order to deal with new classes or even new sub-classes. This task can be performed in real time, and in robust ways due to system's redundancy; **(3)** the present algorithm can work either in unsupervised or supervised mode (adding the final $k$-NNR classification and using training samples as markers), and finally **(4)** The self-organizing nature of ANTIDS makes them an ideal candidate for ID systems. The present method shows how stigmergy can easily be made operational. Indeed, it is a promising first step to design groups of artificial agents which solve problems: replacing coordination (and possible some hierarchy) through direct communications by indirect interactions is appealing if one wishes to design simple agents and reduce communication among agents, in order to implant them on-line in complex large scale computer network infrastructures.

## ACKNOWLEDGEMENT

Thanks to *Joaquim Castelo Ramos* for his time on running part of several *ANTIDS* tests.